\documentclass[aps,prl,twocolumn,showpacs,showkeys,superscriptaddress]{revtex4-1}
\usepackage{graphicx}% Include figure files

\newcommand{\met}{\mbox{${\not\!\!E_{\rm T}}$}}
\newcommand{\metvec}{\mbox{${\not\!\! \vec{E}_{\rm T}}$}}

\begin{document}

%\preprint{}

\title{First Search for Multijet Resonances in \boldmath{$\sqrt{s} = 1.96$}
TeV \boldmath{$ p\bar{p}$} Collisions}




%\author{The CDF Collaboration}
%\thanks{}
\affiliation{}

%\input{March2010_Authors}

\affiliation{Institute of Physics, Academia Sinica, Taipei, Taiwan 11529, Republic of China} 
\affiliation{Argonne National Laboratory, Argonne, Illinois 60439, USA} 
\affiliation{University of Athens, 157 71 Athens, Greece} 
\affiliation{Institut de Fisica d'Altes Energies, ICREA, Universitat Autonoma de Barcelona, E-08193, Bellaterra (Barcelona), Spain} 
\affiliation{Baylor University, Waco, Texas 76798, USA} 
\affiliation{Istituto Nazionale di Fisica Nucleare Bologna, $^z$University of Bologna, I-40127 Bologna, Italy} 
\affiliation{University of California, Davis, Davis, California 95616, USA} 
\affiliation{University of California, Los Angeles, Los Angeles, California 90024, USA} 
\affiliation{Instituto de Fisica de Cantabria, CSIC-University of Cantabria, 39005 Santander, Spain} 
\affiliation{Carnegie Mellon University, Pittsburgh, Pennsylvania 15213, USA} 
\affiliation{Enrico Fermi Institute, University of Chicago, Chicago, Illinois 60637, USA}
\affiliation{Comenius University, 842 48 Bratislava, Slovakia; Institute of Experimental Physics, 040 01 Kosice, Slovakia} 
\affiliation{Joint Institute for Nuclear Research, RU-141980 Dubna, Russia} 
\affiliation{Duke University, Durham, North Carolina 27708, USA} 
\affiliation{Fermi National Accelerator Laboratory, Batavia, Illinois 60510, USA} 
\affiliation{University of Florida, Gainesville, Florida 32611, USA} 
\affiliation{Laboratori Nazionali di Frascati, Istituto Nazionale di Fisica Nucleare, I-00044 Frascati, Italy} 
\affiliation{University of Geneva, CH-1211 Geneva 4, Switzerland} 
\affiliation{Glasgow University, Glasgow G12 8QQ, United Kingdom} 
\affiliation{Harvard University, Cambridge, Massachusetts 02138, USA} 
\affiliation{Division of High Energy Physics, Department of Physics, University of Helsinki and Helsinki Institute of Physics, FIN-00014, Helsinki, Finland} 
\affiliation{University of Illinois, Urbana, Illinois 61801, USA} 
\affiliation{The Johns Hopkins University, Baltimore, Maryland 21218, USA} 
\affiliation{Institut f\"{u}r Experimentelle Kernphysik, Karlsruhe Institute of Technology, D-76131 Karlsruhe, Germany} 
\affiliation{Center for High Energy Physics: Kyungpook National University, Daegu 702-701, Korea; Seoul National University, Seoul 151-742, Korea; Sungkyunkwan University, Suwon 440-746, Korea; Korea Institute of Science and Technology Information, Daejeon 305-806, Korea; Chonnam National University, Gwangju 500-757, Korea; Chonbuk National University, Jeonju 561-756, Korea} 
\affiliation{Ernest Orlando Lawrence Berkeley National Laboratory, Berkeley, California 94720, USA} 
\affiliation{University of Liverpool, Liverpool L69 7ZE, United Kingdom} 
\affiliation{University College London, London WC1E 6BT, United Kingdom} 
\affiliation{Centro de Investigaciones Energeticas Medioambientales y Tecnologicas, E-28040 Madrid, Spain} 
\affiliation{Massachusetts Institute of Technology, Cambridge, Massachusetts 02139, USA} 
\affiliation{Institute of Particle Physics: McGill University, Montr\'{e}al, Qu\'{e}bec, Canada H3A~2T8; Simon Fraser University, Burnaby, British Columbia, Canada V5A~1S6; University of Toronto, Toronto, Ontario, Canada M5S~1A7; and TRIUMF, Vancouver, British Columbia, Canada V6T~2A3} 
\affiliation{University of Michigan, Ann Arbor, Michigan 48109, USA} 
\affiliation{Michigan State University, East Lansing, Michigan 48824, USA}
\affiliation{Institution for Theoretical and Experimental Physics, ITEP, Moscow 117259, Russia}
\affiliation{University of New Mexico, Albuquerque, New Mexico 87131, USA} 
\affiliation{Northwestern University, Evanston, Illinois 60208, USA} 
\affiliation{The Ohio State University, Columbus, Ohio 43210, USA} 
\affiliation{Okayama University, Okayama 700-8530, Japan} 
\affiliation{Osaka City University, Osaka 588, Japan} 
\affiliation{University of Oxford, Oxford OX1 3RH, United Kingdom} 
\affiliation{Istituto Nazionale di Fisica Nucleare, Sezione di Padova-Trento, $^{aa}$University of Padova, I-35131 Padova, Italy} 
\affiliation{LPNHE, Universite Pierre et Marie Curie/IN2P3-CNRS, UMR7585, Paris, F-75252 France} 
\affiliation{University of Pennsylvania, Philadelphia, Pennsylvania 19104, USA}
\affiliation{Istituto Nazionale di Fisica Nucleare Pisa, $^{bb}$University of Pisa, $^{cc}$University of Siena and $^{dd}$Scuola Normale Superiore, I-56127 Pisa, Italy} 
\affiliation{University of Pittsburgh, Pittsburgh, Pennsylvania 15260, USA} 
\affiliation{Purdue University, West Lafayette, Indiana 47907, USA} 
\affiliation{University of Rochester, Rochester, New York 14627, USA} 
\affiliation{The Rockefeller University, New York, New York 10065, USA} 
\affiliation{Istituto Nazionale di Fisica Nucleare, Sezione di Roma 1, $^{ee}$Sapienza Universit\`{a} di Roma, I-00185 Roma, Italy} 

\affiliation{Rutgers University, Piscataway, New Jersey 08855, USA} 
\affiliation{Texas A\&M University, College Station, Texas 77843, USA} 
\affiliation{Istituto Nazionale di Fisica Nucleare Trieste/Udine, I-34100 Trieste, $^{ff}$University of Trieste/Udine, I-33100 Udine, Italy} 
\affiliation{University of Tsukuba, Tsukuba, Ibaraki 305, Japan} 
\affiliation{Tufts University, Medford, Massachusetts 02155, USA} 
\affiliation{University of Virginia, Charlottesville, VA  22906, USA}
\affiliation{Waseda University, Tokyo 169, Japan} 
\affiliation{Wayne State University, Detroit, Michigan 48201, USA} 
\affiliation{University of Wisconsin, Madison, Wisconsin 53706, USA} 
\affiliation{Yale University, New Haven, Connecticut 06520, USA} 
\author{T.~Aaltonen}
\affiliation{Division of High Energy Physics, Department of Physics, University of Helsinki and Helsinki Institute of Physics, FIN-00014, Helsinki, Finland}
\author{B.~\'{A}lvarez~Gonz\'{a}lez$^v$}
\affiliation{Instituto de Fisica de Cantabria, CSIC-University of Cantabria, 39005 Santander, Spain}
\author{S.~Amerio}
\affiliation{Istituto Nazionale di Fisica Nucleare, Sezione di Padova-Trento, $^{aa}$University of Padova, I-35131 Padova, Italy} 

\author{D.~Amidei}
\affiliation{University of Michigan, Ann Arbor, Michigan 48109, USA}
\author{A.~Anastassov}
\affiliation{Northwestern University, Evanston, Illinois 60208, USA}
\author{A.~Annovi}
\affiliation{Laboratori Nazionali di Frascati, Istituto Nazionale di Fisica Nucleare, I-00044 Frascati, Italy}
\author{J.~Antos}
\affiliation{Comenius University, 842 48 Bratislava, Slovakia; Institute of Experimental Physics, 040 01 Kosice, Slovakia}
\author{G.~Apollinari}
\affiliation{Fermi National Accelerator Laboratory, Batavia, Illinois 60510, USA}
\author{J.A.~Appel}
\affiliation{Fermi National Accelerator Laboratory, Batavia, Illinois 60510, USA}
\author{A.~Apresyan}
\affiliation{Purdue University, West Lafayette, Indiana 47907, USA}
% ...
\collaboration{CDF Collaboration\footnote{With visitors from $^a$University of Massachusetts Amherst, Amherst, Massachusetts 01003,
$^b$Istituto Nazionale di Fisica Nucleare, Sezione di Cagliari, 09042 Monserrato (Cagliari), Italy,
$^c$University of California Irvine, Irvine, CA  92697, 
$^d$University of California Santa Barbara, Santa Barbara, CA 93106
$^e$University of California Santa Cruz, Santa Cruz, CA  95064,
$^f$CERN,CH-1211 Geneva, Switzerland,
$^g$Cornell University, Ithaca, NY  14853, 
$^h$University of Cyprus, Nicosia CY-1678, Cyprus, 
$^i$University College Dublin, Dublin 4, Ireland,
$^j$University of Fukui, Fukui City, Fukui Prefecture, Japan 910-0017,
$^k$Universidad Iberoamericana, Mexico D.F., Mexico,
$^l$Iowa State University, Ames, IA  50011,
$^m$University of Iowa, Iowa City, IA  52242,
$^n$Kinki University, Higashi-Osaka City, Japan 577-8502,
$^o$Kansas State University, Manhattan, KS 66506,
$^p$University of Manchester, Manchester M13 9PL, England,
$^q$Queen Mary, University of London, London, E1 4NS, England,
$^r$Muons, Inc., Batavia, IL 60510,
$^s$Nagasaki Institute of Applied Science, Nagasaki, Japan, 
$^t$National Research Nuclear University, Moscow, Russia,
$^u$University of Notre Dame, Notre Dame, IN 46556,
$^v$Universidad de Oviedo, E-33007 Oviedo, Spain, 
$^w$Texas Tech University, Lubbock, TX  79609, 
$^x$Universidad Tecnica Federico Santa Maria, 110v Valparaiso, Chile,
$^y$Yarmouk University, Irbid 211-63, Jordan,
$^{gg}$On leave from J.~Stefan Institute, Ljubljana, Slovenia, 
}}
\noaffiliation

\date{\today}

\begin{abstract}
We present the first model independent search for three-jet hadronic
resonances within multijet events in $\sqrt{s} = 1.96$ TeV $p
\bar{p}$ collisions at the Fermilab Tevatron using the CDF II
detector.  Pair production of supersymmetric gluinos and squarks with
hadronic R-parity violating decays is employed as an example of a new
physics benchmark for this signature.  Selection criteria based on the
kinematic properties of an ensemble of jet combinations within each
event help to extract signal from copious QCD background.  Our
background estimates include all-hadronic $t\bar{t}$ decays that have
a signature similar to the signal.  No significant excess outside the
top quark mass window is observed in data with an integrated
luminosity of 3.2 fb$^{-1}$.  We place 95\% confidence level limits on
the production cross section $\sigma(p \bar{p} \rightarrow XX^{'})
\times \rm{BR}$($\widetilde{g}\widetilde{g}$$ \rightarrow 3~{\rm
jet}+3~{\rm jet})$ where $X, X' = \widetilde{g},~\widetilde{q}$, or
$\widetilde{\bar{q}}$, with ~$\widetilde{q}$, $\widetilde{\bar{q}}
\rightarrow \widetilde{g}$ + jet, as a function of gluino mass, in the
range of 77 GeV/$c^2$ to 240 GeV/$c^2$.
\end{abstract}

\pacs{13.85.-t, 11.30.Pb}
%\keywords{}

\maketitle

Most searches for new physics at high energy hadron colliders use
signatures that require leptons, photons, or missing
transverse energy (\met)~\cite{bib:coord} in order to suppress
backgrounds from QCD.  Final states with multijets and \met~have also
been explored ~\cite{bib:cdfsusy,bib:d0susy}.

In this letter, we present a first new physics search in an entirely
hadronic channel with no \met~signature using data collected with the
Collider Detector at Fermilab (CDF). This data set corresponds to an
integrated luminosity of 3.2 fb$^{-1}$ of $p\bar{p}$ collisions at
$\sqrt{s}=1.96$ TeV at the Tevatron collider.  The search utilizes a
novel approach~\cite{bib:rouven,bib:claudia}: an ensemble of all
possible jet triplets within an event consisting of at least six jets
is used to extract a signal from the multijet QCD backgrounds.  We
model the possible new physics origin for this signature with pair
production of $SU(3)_C$ adjoint Majorana fermions each one decaying
into three quarks~\cite{bib:gluinoref1,bib:gluinoref2}.  This search
is sensitive to models such as hadronic R-parity violating
supersymmetry (RPV SUSY)~\cite{bib:rouven} with a gluino, chargino, or
neutralino lightest superpartner, as well as the hadronic decay modes
of pairs of top quarks or fourth generation quarks and it complements
existing di-jet resonances searches at hadron colliders. Moreover, it
does not require any $b$-quark jet identification which is an
important tool, often used for top quark identification.

The CDF II detector is a multi-purpose particle detector consisting of
tracking and calorimeter systems~\cite{bib:detector}.  The data were
collected using an online event selection that requires at least four
calorimeter jets~\cite{bib:trigeff} with uncorrected transverse energy
$E_T>$ 15 GeV.  A jet is formed by a cluster of calorimeter towers and
reconstructed with a cone algorithm using a fixed cone of $\Delta
R=0.4$~\cite{bib:jetclu}, with $\Delta R=\sqrt{\Delta \eta^2+\Delta
\phi^2}$~\cite{bib:coord}.  In the online selection an additional
request is made for the sum of the transverse energy of all clusters
to be larger than 175 GeV.  At the analysis level, jet energies are
corrected to account for effects such as non-linearities in the
detector response and multiple $p\bar{p}$ collisions in an
event~\cite{bib:jetnim}.

Events are selected with at least six jets with transverse momentum
($p_T$) greater than 15 GeV/$c$ and $|\eta| < 2.5$.  The scalar sum of
the most energetic six jets' $p_T$, $\sum p_T$, is required to be
greater than 250 GeV/$c$ and events with $\met > 50$ GeV are removed.
Multiple interactions, resulting in the reconstruction of more than
one primary vertex in the same event, contribute to the multijet
background.  We require at least one primary vertex and discard events
with more than four primary vertices.  To further reduce this
background, we require jets in an event to originate from near the
same point on the beamline.  We associate tracks with each jet where
possible~\cite{bib:jettracks} by requiring $\Delta R$ between the
track and the jet to be less than 0.4. The mean $z$-coordinate of all
tracks associated with each jet ($\bar{z_j}$ for the $j^{\rm th}$
jet), and the associated standard deviation ($\delta(z_j)$) are
determined.  Events with jets that have $|\bar{z_j}| > 60$ cm are
discarded.  We then evaluate the standard deviation of the $\bar{z_j}$
of all jets in the event ($\delta (z_\textrm{all}$)) and select events
that have at least four jets with $\delta(z_j)<4$ cm, and $\delta
(z_\textrm{all}) <0.5$ cm, consistent with the resolution of tracks
associated with jets.  Once the selection is applied, pileup effects
are significantly reduced.  Since we select events with at least six
jets, we consider an ensemble of 20 (or more) possible jet triplets.
We discard those triplets that have more than one jet with no $z$
information.  In addition, all jets in the triplet must have
$\delta(z_j)<2.5$ cm, and originate from within 10 cm of the primary
vertex of the event.

\begin{figure}[t]
\includegraphics[width=3.3in,height=2.3in]{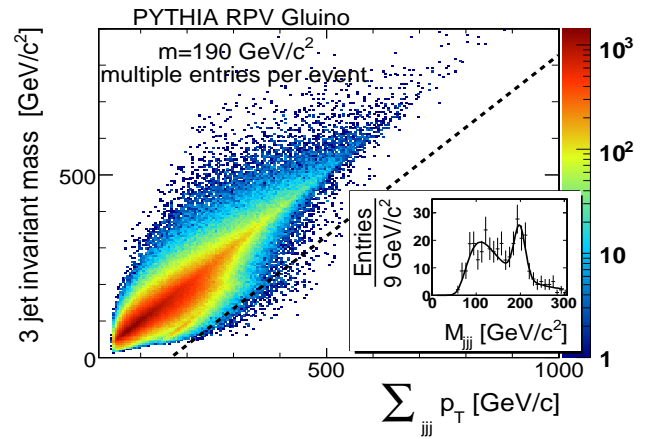}%
\caption{Distribution of $M_{jjj}$ versus $\sum_{jjj}p_T$ for a
  pair-produced RPV gluino with invariant mass 190 GeV/c$^2$ generated
  with {\sc pythia} MC.  Triplets to the right of a diagonal offset
  ($\sum_{jjj} p_T - M_{jjj} c = \Delta$), indicated by the dashed
  line, are are kept.  The inset shows the $M_{jjj}$ distribution for
  the RPV signal MC and with no QCD background after a diagonal offset
  of 195 GeV/$c$ along with a Gaussian plus a Landau fit; the Landau
  shows the combinatorial contribution within the signal jet
  ensemble. The QCD background distribution resembles that of the
  combinatorial contribution, because they are both due to effectively
  uncorrelated triplets.  \label{fig:mass_sumpt}}
\end{figure}

The biggest challenge of this analysis is to reduce the large multijet
QCD background.  To extract signal from this background, we apply the
following technique: for every accepted triplet we calculate the
invariant mass, $M_{jjj}$, and scalar sum $p_T$, $\sum_{jjj}p_T$.
Triplets made of uncorrelated jets tend to have $M_{jjj} c \approx
\sum_{jjj}p_T $, while signal triplets should have $M_{jjj}$ as close
to the mass of the decaying particle as allowed by jet energy
resolution.  We then select triplets with $\sum_{jjj} p_T - M_{jjj} c
> \Delta$, $\Delta$ being a diagonal offset as illustrated in
Fig.~\ref{fig:mass_sumpt}.  The diagonal offset values are optimized
for the best signal over background ratio separately for each hadronic
resonance mass in this search.  The optimized diagonal offset
selection greatly reduces the QCD background and the contribution from
incorrect combinations of jets.  We note that for a small fraction of
events it is possible for multiple triplets to pass all selection
criteria.

The QCD background is estimated from a 5-jet data sample, which is
statistically independent of the signal sample of $\ge 6$ jets (for
brevity referred to as 6-jet).  The 5-jet $M_{jjj}$ distribution is
rescaled by the ratio of the 6-jet to 5-jet population in each
$\sum_{jjj}{p_T}$ bin.  A Landau function is chosen~\cite{bib:rouven}
to fit the scaled 5-jet $M_{jjj}$ distribution.  The Landau parameters
extracted from the scaled 5-jet $M_{jjj}$ distribution vary by less
than 2 GeV/$c^2$ from similar fits to the 6-jet sample, indicating
that the scaled 5-jet sample describes the background in the 6-jet
sample well.  The contribution to the background from $t\bar{t}$ pair
production is estimated using the {\sc pythia} Monte Carlo (MC)
generator~\cite{bib:pythia} followed by the CDF detector
simulation~\cite{bib:cdfsim}.  These events were generated assuming a
top quark mass of 172.5 GeV/$c^2$ and production cross section of 7.5
pb.  To ensure a proper fit to the QCD background, the fit is blinded
to the mass region corresponding to the top quark, 153
GeV/$c^2$$<M_{jjj}<$189 GeV/$c^2$.  Additionally, we find that
truncating the Landau fit for lower values of $\Delta$ gives an
improved description of the QCD background.  The Landau parameters
extracted from the fits vary smoothly as functions of the diagonal
offset value.  We now have a firm prediction for the QCD background
and fix the parameters when we fit for signal.

The signal is modeled using the {\sc pythia} MC generator.  The
process $p \bar{p} \rightarrow XX^{'}$ where $X, X' =
\widetilde{g},~\widetilde{q}$, or $\widetilde{\bar{q}}$ is simulated
at several gluino mass values, ranging from 74~GeV/$c^2$ to
245~GeV/$c^2$ with hadronic $uds$ RPV SUSY decays turned on, allowing
gluino decays to three light jets. Two scenarios of squark masses are
considered ($0.5 $ TeV/$c^2$ $< m_{\widetilde{q}} < 0.7 $ TeV/$c^2$,
$m_{\widetilde{q}} = m_{\widetilde{g}} + 10 $ GeV/$c^2$) and were
found to give equivalent acceptances.

The acceptance of the trigger, reconstruction, and selection
requirements for signal events is determined by fitting the pair
produced RPV gluino MC with a Landau plus Gaussian function,
corresponding to the combinatorial contribution and signal peak
respectively.  An example is shown in the inset of
Fig.~\ref{fig:mass_sumpt}.  The Gaussian is integrated in a $\pm
1\sigma$ range to extract the number of signal triplets.  This procedure
is repeated for various diagonal offset values and the optimal offset
for each hadronic resonance mass is determined.  The acceptance,
calculated for these optimal offset values, is $5 \times 10^{-5}$,
constant within 20\% across all gluino mass points.

The expected sensitivity of this analysis in the absence of signal is
determined with a set of background-only experiments
(pseudoexperiments).  A pseudoexperiment is constructed with the
background modeled by a Landau function whose parameters are chosen
randomly from within the range allowed by the background shape fits,
with the expected amount of $t\bar{t}$ added.  Each pseudoexperiment
is fit with the Landau background shape parameters fixed, and a signal
Gaussian whose position is determined by the mass point being fit, and
whose amplitude and width are allowed to vary within a range
determined by the expected signal shape.  The number of signal triplets
allowed by each pseudoexperiment is extracted by integrating the
Gaussian in the same way as in the acceptance calculation.

Two broad categories of systematic uncertainties are accounted for in
extracting a cross section: uncertainties in the shape of the
$M_{jjj}$ distribution and uncertainties in the acceptance of the
signal.  Shape uncertainties, determined from background and signal
fits, are incorporated in the pseudoexperiments themselves.
Acceptance uncertainties arise from modeling the signal Monte Carlo
and include effects of initial and final state
radiation~\cite{bib:isrfsr} (20\%), parton distribution functions
(PDFs) from CTEQ~\cite{bib:cteq} (10\%), jet energy
scale~\cite{bib:jetnim} (31\%) and luminosity~\cite{bib:lumi} (6\%)
uncertainties.  The overall acceptance uncertainty due to these sources is 38\%.

\begin{figure}[t]
\includegraphics[width=3.4in,height=2.4in]{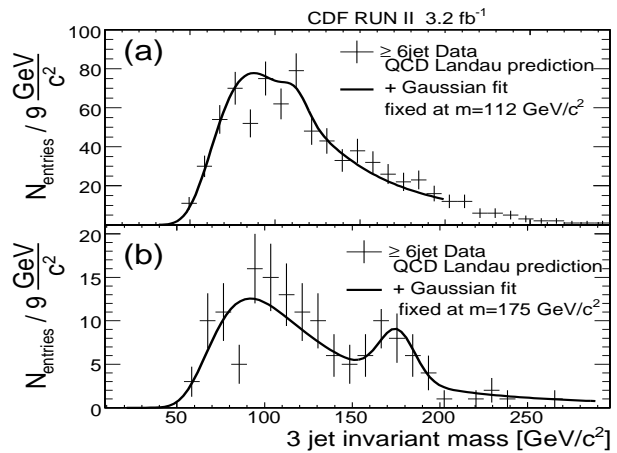}%
\caption{$M_{jjj}$ distributions in 3.2 fb$^{-1}$ data fitted to a
  Landau (parameters are extracted from fits to the scaled 5-jet $M_{jjj}$ distribution) plus a Gaussian at (a) 112 GeV/$c^2$ (optimal diagonal offset  
  value 155 GeV/$c$) and (b) 175 GeV/$c^2$ (optimal diagonal offset value
  of 190 GeV/$c$).  The fit function in panel (b) includes a Gaussian fixed at $m=175$~GeV/$c^2$.
    \label{fig:mass}}
\end{figure}

We search for a hadronic resonance in the data for an invariant mass
($m$) 77 -- 240 GeV/$c^2$ in 9 GeV/$c^2$ steps,
consistent with jet energy resolution.  For each mass, jet triplets
are selected by the optimal diagonal offset value.  The data $M_{jjj}$
distribution is fit in exactly the same way as the pseudoexperiments.
Figure~\ref{fig:mass} shows the $M_{jjj}$ distribution for $m$ = 112
GeV/$c^2$ and 175 GeV/$c^2$.  The latter fit shows a noticeable excess
consistent in mass with the hadronic decay of the top quark.  The
Gaussian component of the fit integrated from 165 GeV/$c^2$ to 185
GeV/$c^2$, corresponding to a $\pm 1 \sigma$ window around the
Gaussian peak, gives $11 \pm 5$ triplets.  The number of expected QCD
background triplets in the same mass window from the Landau function
is $8 \pm 1$.  The $t\bar{t}$ contribution to background is evaluated
using {\sc pythia}.  It is cross-checked with higher order $t\bar{t}$
MC generators  {\sc alpgen}~\cite{bib:alpgen} and {\sc
mc@nlo}~\cite{bib:mcatnlo}, samples that varied the amount of initial
and final state radiation, as well as samples that varied the PDFs
within their uncertainties.  These studies lead us to expect between
0.5 and 1.1 triplets from $t\bar{t}$ production in the aforementioned
mass range.  We note that $\sim$10\% of the triplets in the top mass
window originate from two or more combinations in a jet ensemble of a
given event, consistent with the {\sc pythia} $t\bar{t}$ simulation.
We evaluate the significance of the excess using the pseudoexperiment
method described above, which includes systematic uncertainties on
signal acceptance as well as the shape of the $M_{jjj}$ distribution.
The observed excess is 2 standard deviations (2$\sigma$) above the
prediction. Additional cross-checks, such as requiring one of the jets
to have originated from a $b$-quark, suggest that the excess is
consistent with coming from top quarks.

\begin{figure}[t]
\includegraphics[width=3.4in,height=2.4in]{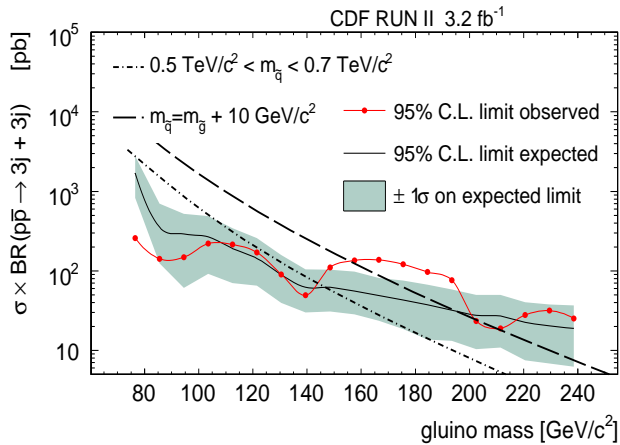}
\caption{The observed (points) and expected (solid black line) 95\%
  confidence level limits on the production cross section
  $\sigma(p \bar{p} \rightarrow XX^{'}) \times \rm{BR}$($\widetilde{g}\widetilde{g} \rightarrow 3~{\rm jet}+3~{\rm jet}$) 
  where
$X, X' = \widetilde{g},~\widetilde{q}$, or $\widetilde{\bar{q}}$,
 including systematic uncertainties.  The shaded bands represent the total uncertainty on the limit. Also shown is the model cross section from {\sc pythia}
  corrected by an NLO $k$-factor (dash-dot line for $0.5 $ TeV/$c^2$ $< m_{\widetilde{q}} < 0.7 $ TeV/$c^2$, dashed line for $m_{\widetilde{q}} = m_{\widetilde{g}} + 10 $ GeV/$c^2$).  
    \label{fig:limit}}
\end{figure}

We do not observe a significant deviation from standard model
backgrounds anywhere in the data.  A Bayesian approach is used to
place 95\% confidence level limits on $\sigma(p \bar{p} \rightarrow
XX^{'}) \times \rm{BR}$($\widetilde{g}\widetilde{g}$$ \rightarrow
3~{\rm jet}+3~{\rm jet})$ where $X, X' =
\widetilde{g},~\widetilde{q}$, or $\widetilde{\bar{q}}$, with
~$\widetilde{q}$, $\widetilde{\bar{q}} \rightarrow \widetilde{g}$ +
jet, versus gluino mass, shown in Fig.~\ref{fig:limit}.  The largest
excess observed is the one previously noted located near the top quark
mass.  We find that our background estimate has a 2.3\% probability of
producing such a deviation.  Comparisons to the theoretical cross
section for $\sigma( p\bar{p} \rightarrow XX^{'} ) \times
\rm{BR}$($\widetilde{g} \widetilde{g}$$ \rightarrow 3~\rm{jet} +
3~\rm{jet})$ from {\sc pythia} corrected by a next-to-leading-order
(NLO) $k$-factor calculated using {\sc prospino}~\cite{bib:Prospino}
are shown in the dashed and dash-dot lines for two different squark
mass scenarios.  For a decoupled squark mass ($0.5 $ TeV/$c^2$ $<
m_{\widetilde{q}} < 0.7 $ TeV/$c^2$) we exclude gluinos below a mass
of 144 GeV/$c^2$ (dashed line).  In the case of a squark mass which is
nearly degenerate with the gluino mass ($m_{\widetilde{q}} =
m_{\widetilde{g}} + 10 $ GeV/$c^2$) we exclude gluinos below 155
GeV/$c^2$ (dash-dot line).

We have performed a first search for three-jet hadronic resonances in
a six or more jet final state using a data sample with an integrated
luminosity of 3.2 fb$^{-1}$ collected by the CDF II detector.  A novel
technique is introduced that exploits kinematic features within an
ensemble of jet combinations that allows us to extract signal from the
QCD background.  We observe no significant excess in the data in an
invariant mass range from 77 GeV/$c^2$ to 240 GeV/$c^2$ and place 95\%
confidence level limits on the production cross section $\sigma(p
\bar{p} \rightarrow XX^{'}) \times
\rm{BR}$($\widetilde{g}\widetilde{g}$$ \rightarrow 3~{\rm jet}+3~{\rm
jet})$ where $X, X' = \widetilde{g},~\widetilde{q}$, or
$\widetilde{\bar{q}}$, with ~$\widetilde{q}$, $\widetilde{\bar{q}}
\rightarrow \widetilde{g}$ + jet, versus gluino mass.  The results are
presented as limits on RPV gluinos decaying to three jets, but
are more widely applicable to any new particle with a
three-jet decay mode.  Two different squark mass scenarios have been
considered: decoupled squarks and squarks nearly degenerate in mass
with the gluino.  We can exclude gluinos below 144 GeV/$c^2$ and
155 GeV/$c^2$ respectively.

\begin{acknowledgments}
We thank R. Essig, S. Mrenna, M. Park, and Y. Zhao for
assistance with this analysis. We also thank the Fermilab staff and
the technical staffs of the participating institutions for their vital
contributions. This work was supported by the U.S. Department of
Energy and National Science Foundation; the Italian Istituto Nazionale
di Fisica Nucleare; the Ministry of Education, Culture, Sports,
Science and Technology of Japan; the Natural Sciences and Engineering
Research Council of Canada; the National Science Council of the
Republic of China; the Swiss National Science Foundation; the
A.P. Sloan Foundation; the Bundesministerium f\"ur Bildung und
Forschung, Germany; the Korean World Class University Program, the
National Research Foundation of Korea; the Science and Technology
Facilities Council and the Royal Society, UK; the Institut National de
Physique Nucleaire et Physique des Particules/CNRS; the Russian
Foundation for Basic Research; the Ministerio de Ciencia e
Innovaci\'{o}n, and Programa Consolider-Ingenio 2010, Spain; the
Slovak R\&D Agency; and the Academy of Finland.
\end{acknowledgments}

\bibliography{cdfmultijets_prl}

\begin{thebibliography}{20}
\bibitem [{bib({\natexlab{a}})}]{bib:coord}%
  \BibitemOpen
  \bibinfo {note} {{C}DF uses a ($z$, $\phi$, $\theta$) coordinate system with
  the $z$-axis in the direction of the proton beam; $\phi$ and $\theta$ are the
  azimuthal and polar angle respectively. The pseudorapidity is defined as
  $\eta = -\ln(\tan{\theta \over 2})$, and the transverse momentum and energy
  as $p_{\rm T} = p \sin \theta$ and $E_{\rm T} = E \sin \theta$, respectively.
  Missing transverse energy ($\met = |\metvec|$) is defined as $\metvec = -
  \sum_{i} E_{\rm T}^i {\bf{\hat{n}_i}}$ where ${\bf{\hat{n}_i}}$ is a unit
  vector in the transverse plane that points from the beam-line to the $i^{th}$
  calorimeter tower.}\BibitemShut {Stop}%
\bibitem [{bib({\natexlab{b}})}]{bib:jettracks}%
  \BibitemOpen
  \bibinfo {note} {{N}ote that tracking efficiency drops significantly beyond
  $|\eta|>1.0$. Jets at large $|\eta|$ do not have associated tracks and have
  only limited $z$ information.}\BibitemShut {Stop}%
\end{thebibliography}

\end{document}